\documentclass[article]{aa}
\usepackage{rotating}
\usepackage{longtable}
\usepackage{subfigure}
\usepackage{amssymb}
 \usepackage{graphicx}
 \usepackage{astro_bib_macro}
\usepackage[authoryear]{natbib}

\begin{document}
 
\title{HD~97048 : a closer look to the disk
\thanks{Based on observations obtained at VLT (Paranal) with VISIR. Program number 075.C-0540(C)}
 }
 \author{C. Doucet\inst{1,2} \and E. Habart\inst{3} \and E. Pantin\inst{1,2} \and C. Dullemond\inst{4} \and P. O. Lagage\inst{1,2} \and C. Pinte\inst{5} \and G.Duch\^ene\inst{5} \and F. M\'enard\inst{5}} 
 
 \institute{AIM, Unit\'e Mixte de Recherche CEA - CNRS - Universit\'e Paris VII -UMR 7158 ,France \and  DSM/DAPNIA/service d'Astrophysique, CEA/Saclay, F-91191 Gif-sur-Yvette, France \and Institut d'Astrophysique Spatiale (IAS), 91405 Orsay cedex, France \and Max-Planck-Institut fur Astronomie Heidelberg, Konigstuhl 17, Heidelberg, Germany \and Laboratoire d'Astrophysique de Grenoble, CNRS/UJF UMR 5571, Grenoble, France}
 
 \offprints{C.Doucet,
 \email{coralie.doucet@gmail.com}}

 \abstract{}
 {Today, large ground-based instruments, like VISIR on the VLT,
providing diffraction-limited ($\sim$0.3 arcsec) images in the mid-infrared where strong PAH features appear enable us to see the flaring structure of the disks around Herbig Ae stars. 
 Although great progress has been made in modelling the disk with radiative transfer models able to reproduce the spectral energy distribution (SED) of Herbig Ae stars, the constraints brought by images have not been yet fully exploited. Here, we are interested in checking if these new observational imaging constraints
can be accounted for by predictions based on existing models of passive centrally irradiated hydrostatic disks made to fit the SEDs of the Herbig Ae stars.}
{The images taken by VISIR in the 8.6 and 11.3 $\mu$m aromatic features reveal a large flaring disk around HD97048 inclined to the line of sight. In order to analyse the spatial distribution of these data, we use a disk model which includes the most up to date understanding of disk structure and physics around Herbig Ae stars with grains in thermal equilibrium in addition to transiently-heated PAHs. }
 { We compare the observed spatial distribution of the PAH emission
feature and the adjacent continuum emission with predictions based on existing full disk models. 
 Both SED and spatial distribution are in very good agreement with the model predictions for common disk parameters.
 }{ We take the general agreement between observations and predictions as a strong support for the physical pictures underlying our flared disk model.}
 

  \maketitle
 
 \section{Introduction}
 
 Herbig Ae/Be stars are thought to be the intermediate mass ($\sim$ 2-10 M$_{\odot}$) counterparts of T Tauri stars \citep{WAWA98}. They have a strong infrared (IR) excess coming from warm circumstellar dust. Similar to T Tauri stars, this dust is believed, due to the interpretation of spectral energy distributions (SEDs), to reside in a flared/flat circumstellar disk \citep{MEEUS01}. 

PAHs (Polycyclic Aromatic Hydrocarbons) undergo transient heating : they do not reach thermal equilibrium with the radiation field but absorb individual photons, experiencing a rapid increase in temperature, slowly cool, thus re-radiating the absorbed energy in the infrared (IR). This radiation allows to see much further in the disk since these grains can reach high temperatures far away from the star. As a result, PAHs emission can be used to probe the external region of disks around HAe stars \citep{LAGA06}.
Another point of interest is that, if present, PAHs can constitute
an important source of opacity and are likely  
to play a key role in the thermal budget and chemistry of the gas,
as they do in the interstellar medium.
The thermal coupling between PAHs and gas via the photoelectric  
effect will, for example, determine the
gas temperature in the upper layer of the disk where the gas and dust temperatures are not
well coupled (Jonkheid et al. 2004; Kamp \& Dullemond 2004).
Furthermore, PAHs are also good tracers of the presence of very small particles 
in the surface layers of disks
and their emission can tell if and where very small particles survive settling and
coagulation processes that cause the majority of the original grain population
to grow to larger sizes in the same objects.

We have started a program of imaging nearby HAe stars with VISIR, the new mid-infrared instrument attached to the third 8-m unit of ESO's Very Large Telescopes located at Cerro Paranal, Chile. One of the first targets was HD~97048 which has a strong IR excess, characteristic of a flared disk, and strong PAH features.

In a previous work, we were able for the first time to constrain the geometry of the disk, thanks to PAH emission \citep{LAGA06}. \citet{LAGA06} found there is a large flaring disk around HD97048 extending at least up to 370 AU, vertically optically thick at the observed wavelength (8.6 $\mu$m)
and inclined to the line of sight by 42.8$^{+0.8}_{-2.5}$ degree.
Morever, with a very simple model, we were able to measure the flaring index which was found to be 1.26$^{+0.05}_{-0.05}$. In this model, the PAH-emitting
region is only located at the surface of the disk,
whose surface scale height H$_{s}$ varies with r$^{q}$ ($q$ is the flaring index) and whose flux intensity I follows r$^{-p}$.
It is only recently that images of the flaring structure of the disk around Herbig Ae objects became available \citep{PERRIN06, LAGA06}. It is now primordial to confront the spatial distribution constraints with existing full disk models only based on fitting the SEDs of the Herbig Ae stars.

In this paper, we are interested in particular to check if these new observational spatial constraints
can be accounted for by predictions based on existing models of passive centrally irradiated hydrostatic disks made to fit the SEDs of the Herbig Ae stars.  We explore whether a more sophisticated disk model could explain the SED of HD 97048 and at the same time, the spatial distribution of the circumstellar material with little free parameters. 
To the best of our knowledge, few tests like that \citep{DOUCET06a} have been done yet
for disks around pre-main sequence stars of intermediate mass. 
To do that, we used the model of \citet{DDN01,DD04a} which could already account for the global shape of the SED of a quite large number of Herbig Ae stars.
The dust model takes into account grains at thermal equilibrium and stochastically-heated PAH.
The structure of the disk is calculated with hydrostatic equilibrium and a radiative transfer in two dimensions is used to calculate the emission of the different grains population.

The paper is organized as follows. In Sect. 2, we present the knowledge on HD 97048 and its circumstellar material. In Sect. 3, we describe the observations and the data results. 
In Sect. 4, we describe the disk model and in Sect. 5, we compare the observed SED and spatial distribution of the circumstellar material to the model predictions.
\begin{table*}[!t]
 \begin{center}
 \caption{Observations of HD~97048 in the different VISIR filters in imaging BURST mode. The sensitivity were calculated for each night in the BURST mode. The filter are free of any strong atmospheric line contribution.}
 \begin{minipage}{\hsize}
\setlength{\tabcolsep}{1.2mm}
\renewcommand{\arraystretch}{1.3}
\renewcommand{\footnoterule}{}
  \begin{tabular}{lccccccccc}
 \hline
    Filter & Central   & half band width &  Sensitivity & Elementary  & Total time &  Date & Seeing & Airmass & Standard  \\
          &wavelength  &                 &              & exposure time     & on source  &    &     &        &   star    \\
	  &  ($\mu$m)            & ($\mu$m)   &(mJy/10$\sigma$/1h)&        (ms)         & (s)  &     & (arcsec) &    &        \\
  \hline
  \hline
  SIV & 10.49  & 0.16 & 5.6 &  25& 600 & 17/06/2005 & 0.80 & 1.74 & HD91056 \\
  PAH2 & 11.26 & 0.59 & 4.5 & 50 & 800 & 25/01/2005 & 0.55 & 1.66 & HD85503\\
  NeII & 12.27 & 0.18 & 7.7 &16& 320 & 17/06/2005 & 0.80 & 1.74 & HD91056 \\
  \hline
 \end{tabular}\label{filter}
 \end{minipage}
  \end{center}
 \end{table*}

\section {HD97048}

HD97048 is a nearby HAe star of spectral type Be9.5/A0 located in the Chamaeleon I dark cloud, at a distance of 180  pc (\citet{VDA97b}, \citet{WHI97}).
It is surrounded by a large amount of circumstellar material left from the star formation process, which produces a large infrared (IR) excess over the stellar emission (large IR$\sim$0.35-0.40L$_{\star}$, \citet{ACKE04b,VKK02}). 
HD 97048 has been classified as a HAe star of group I with evidence of flared disk \citep{MEEUS01},
since its SED is rising in the IR \citep{ACKE04b}. 
Spectroscopic observations of the IR excess have also revealed the presence of strong PAH features at 3.3, 6.2, 7.7, 8.6, 11.3 microns and nano-diamonds features in the 3.4-3.5 $\mu$m region \citep{SIEBEN00,VKK02}. No silicate emission band
at 10 $\mu$m appears in the spectra of HD 97048.
Recent mid-IR long slit spectroscopic observations with TIMMI2 show that the aromatic emission features at long wavelength (i.e., 8.6 and 11.3 $\mu$m) are extended and come mostly from a region of 200-300 AU, likely a disk \citep{VB04}. These results have been confirmed recently by imaging data taken in the PAH band at 8.6 $\mu$m \citep{LAGA06}.  
The large extended part of the mid-IR emission seen on scales of 5 to 10 arcsec by \citet{PRUSTI94} and \citet{SIEBEN00} is most likely due to an extended envelope of transiently heated very small grains and PAHs surrounding the star and the disk system.
Using Adaptive Optics high angular resolution ($\sim$0.1 arcsec) spectroscopic observations, Habart et al. (2004, 2005) were also 
able to spatially resolve the emission in the aromatic and diamond features around 3 $\mu$m and found that the emission must be within 30 AU closely related to the star-disk system.
Finally,  \citet{WEIN05} have recently reported near-IR molecular hydrogen emission for this object.\\


 \section{VISIR observations}\label{observations_visir}
\subsection{Observations}
 The observations were performed using the ESO mid-infrared instrument VISIR installed on the VLT (Paranal, Chili), equiped with a DRS (former Boeing) 256$\times$256 pixels BIB detector. The object was observed on 2005, January 25$^{th}$ and June 17$^{th}$. It was observed in the PAH bands at 8.6 and 11.3 $\mu$m and in the adjacent continuum. The data at 8.6 $\mu$m were already used in \citet{LAGA06}.
In this paper, we focus on the data obtained at 11.3 $\mu$m. A summary of the observations is given in Table ~\ref{filter}. These data were taken with an imaging mode of VISIR which allows diffraction-limited image in the N band: the BURST mode. 
 


Under good seeing ($\leq$ 0.5 arcsec in the visible), the images in the mid-IR are diffraction-limited even on a 8 meter class telescope. Unfortunately, the median seeing experienced at Paranal is rather of 0.8 arcsec, which degrades the angular resolution. Indeed, for a seeing of 0.8 arcsec in the visible, assuming that the wavelength dependence of the seeing follows a $\lambda^{-1/5}$ law, the seeing value at 10 $\mu$m is 0.4 arcsec, which is larger than the diffraction limit of 0.3 arcsec. This represents a 5 pixels movement on the detector with the smallest field of view of VISIR (0.075"/pixel). In order to reach the best
spatial resolution with VISIR, we implemented a new imaging mode on bright objects, the \emph{BURST mode}. The
principle is to take short enough exposure images ($\lesssim$ 50 ms) to freeze the turbulence; the coherence time of the atmosphere at 10 $\mu$m is around 300 ms at Paranal for a good seeing. In order to correct for the turbulence by offline processing, the data are stored every 1000 elementary images in one nodding position for a chopping frequency of 0.25 Hz in the direction north/south. The nodding direction is perpendicular to the chopping direction with an amplitude of 8". After classical data reduction in mid-IR, a cube of 500\footnote{1000 divided by 2 because of the 2 chopper positions} images chopped and nodded (4 beams/image) is obtained. Because of the turbulence, each source on an image moves independently and as a result, we have to extract individually the 4 sources in each image (4 quarters) of the cube and shift and add the image with the ones corresponding to the same quarter. Finally, we shift and add the four final images of the four quarters.\\

  \begin{figure}[!h]
\centerline{\resizebox{8.55cm}{!}{\includegraphics[angle=0,scale=1.]{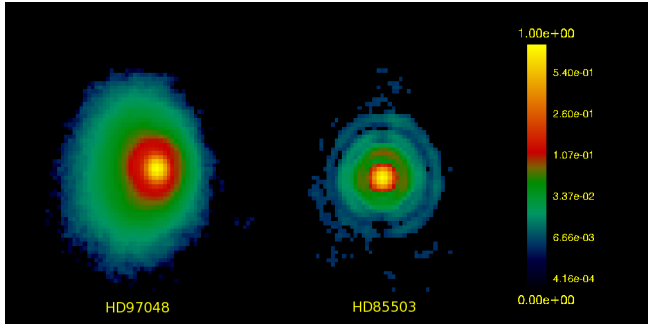}}}
\caption{\emph{On the left}, HD~97048 in the PAH2 filter (centered at 11.3 $\mu$m) with VISIR camera.\emph{ On the right}, PSF reference (HD85503) with the same filter. The two object are normalized in order to see the extension. The object is more extended than the reference with an asymmetry east/west in the emission.}\label{ima}
\end{figure}

 \subsection{Results}
 
In this section, we present the images of HD~97048 obtained in the PAH
emission filter, as well as in the adjacent continuum emission, 
and compare it with the standard star in the same filters.

\begin{figure}[!h]
\centerline{\includegraphics[angle=0,scale=0.35]{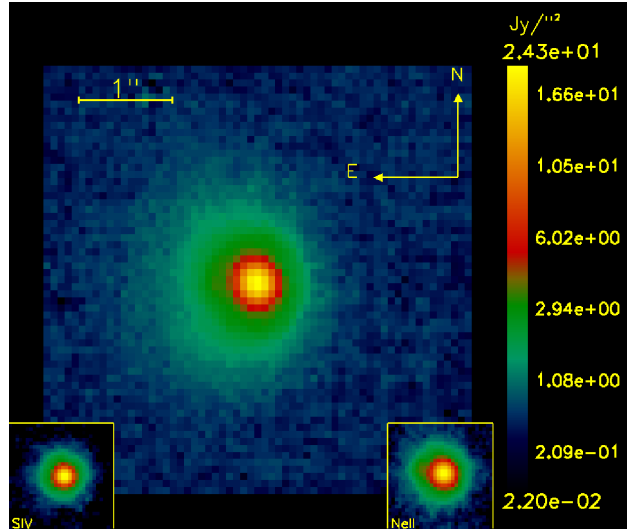}}

\caption{HD~97048 in the PAH2 filter (\emph{image in the middle}), and in the adjacent continuum (SIV \emph{on the left}, NeII \emph{on the right}). The images have the same signal-to-noise so that it is possible to compare the extension.}\label{HD97048_pah2_continuum}

\end{figure}

\begin{figure}[!h]
\centerline{\resizebox{8.55cm}{!}{\includegraphics[angle=0,scale=1.]{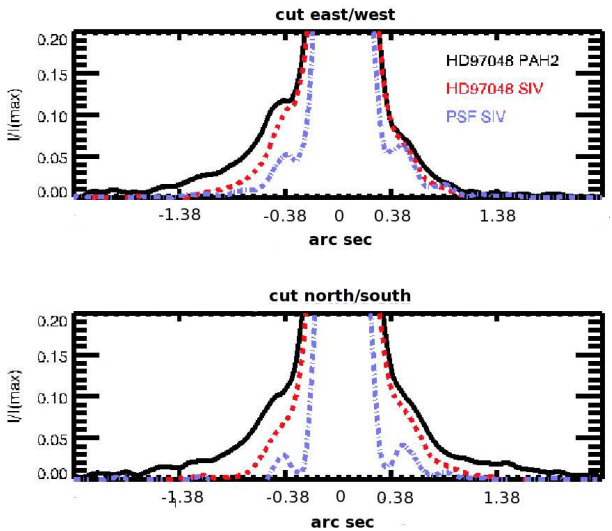}}}
\caption{Normalized intensity profiles along a cut through the VISIR images of HD~97048 in the PAH band (full line) and in the continuum SIV (dashed line) superimposed on a reference star (dashed-dot line) in SIV band. Direction East/West is in the upper panel and above, it represents the direction South-North.
 Comparing the PAH band and the adjacent continuum (for a same signal-to-noise), HD~97048 is much more extended in the PAH band. Comparing the PSF and the object in SIV, we see that HD~97048 is also extended in the continuum.}\label{oplot_pah2_s4}
 \end{figure}

Fig.~\ref{ima} shows the image of HD~97048 in the PAH filter centered at 11.3 $\mu$m and compared with the standard star HD85503.
HD~97048 is quite extended - up to 2 arcsec - compared to the reference star. 
The full-width at half-maximum (FWHM) of HD~97048 is 1.2 times that of the standard star, close to the diffraction-limit, for almost all filters (Tab.~\ref{resultat_obs}).
In addition, comparing the emission in the PAH band and in the adjacent continuum (Fig.~\ref{HD97048_pah2_continuum}), the emission in PAH is much more extended than in the continuum. Fig.~\ref{oplot_pah2_s4} shows the brightness spatial distribution along a cut in the north/south and east/west direction in the PAH filter and in the adjacent continuum (SIV) compared to the standard star in SIV filter. 
In the PAH filter, the disk is extended in the direction north/south up to 320 AU (surface brightness of 53 mJy/"$^{2}$) whereas the extension in the continuum at 10.5 $\mu$m goes to 135 AU (surface brightnesss of 110 mJy/"$^{2}$). These results are in agreement with those already found in the previous study at 8.6 $\mu$m \citep{LAGA06}. Furthermore, it is also interesting to point out that the continuum is extended, when comparing the object and the reference star in the SIV filter. 
Finally, by comparing the VISIR and ISO (Infrared Space Observatory) fluxes, we found that the PAH emission peak to the continuum is stronger in the ISO spectrum. This is certainly due to the surrounding nebula, also included in the larger beam of ISO, as already suggested by \citet{VB04}.
Based on the fluxes measured in the wavelength bands at 8.6 and 11.3 $\mu$m, we estimate that the nebula contamination is about 40\%,
in agreement with that found by \citet{VB04}.


\begin{table*}[!ht]
  \begin{minipage}{\hsize}
\setlength{\tabcolsep}{0.75mm}
\renewcommand{\arraystretch}{1.5}
\renewcommand{\footnoterule}{}
 \begin{center}
 \caption{Comparaison of the FWHM for the object and the reference star (PSF) in the different filters. We also mention the theorical value of the FWHM in order to show that the diffraction-limit is obtained with the BURST mode. In order to see the spatial extension, we note in the third and fourth columns the distance from the star at 10 $\sigma$ above the noise in the east/west direction. }
  \begin{tabular}{lc|ccccc|cc|c}
 \hline
           &                   &  \multicolumn{5}{|c|}{HD97048}                & \multicolumn{2}{c|}{PSF}     &           \\  
	          
    Filter & Central  &  Flux                                          &      East   & West   & FWHM  & FWHM   & FWHM & FWHM & FWHM    \\                                             
	  & wavelength   & measured  & (10 $\sigma$) & (10 $\sigma$)& (east/west) &(north/south) & (east/west) & (north/south) & (diffraction) \\
	  &  ($\mu$m)    & (Jy)                               & (arcsec)    &  (arcsec)         & (mas)  &(mas) & (mas) & (mas)  &(mas)  \\
  \hline
  \hline
 
 SIV & 10.49 &4.2 $\pm$ 0.1&  0.67     &0.60  & 322 $\pm$ 15 & 337 $\pm$ 15 & 277 $\pm$ 15 & 262 $\pm$ 15& 262 \\
 PAH2 & 11.26&7.7 $\pm$ 0.1& 1.60 & 0.90 & 337 $\pm$ 15 & 360 $\pm$ 15 & 300 $\pm$ 15 & 285 $\pm$ 15& 285\\
 NeII & 12.27& 6.7 $\pm$ 0.2 &0.67    & 0.60   & 390 $\pm$ 15 & 412 $\pm$ 15 & 300 $\pm$ 15 & 307 $\pm$ 15& 307    \\
 \hline
 \end{tabular}\label{resultat_obs}
\end{center}
 \end{minipage}
 \end{table*}

 \section{Disk model}\label{disk_model}

In order to analyse the observations, we use the disk model described in \citet{DDN01,DD03,DD04a},
including the most up to date understanding of disk structure and physics around Herbig Ae/Be stars.
In the following, we describe the disk structure and the radiation transfer used and give the details of the adopted dust model.
 
 \subsection{Disk structure}
  We consider disks heated by irradiation from the central star, in hydrostatic equilibrium in the vertical direction, with dust and gas well mixed (flared disks, \citet{CG97}). 
  In this model, the stellar flux impinging with flaring angle $\alpha$ upon the disk is absorbed in the upper layers of the disk, which will reradiate half of the flux away from the disk and half down into its deeper layers. The inner boundary (rim) is directly exposed to the stellar flux and is puffed up since it is hotter than the rest of the disk. The model computes how much the inner rim puffs up, and how much of the disk behind it will be shadowed by this puffed-up rim \citep{DDN01}.
 Once the star is given, the disk structure (i.e pressure scale height and flaring angle) is completely defined after specifying the inner and outer radii, the surface density distribution ($\Sigma=\Sigma_{0} (R/R_{0})^{-p}$,with R$_{0}$ a fiducial radius) and the dust model. The models are appropriate for disks that are optically thick to the stellar radiation. It is the case for disks around pre-main sequence stars, up to very large radii (e.g., 5000 AU for a disk mass M$_{d} \sim 0.2 M_{\odot}$, p=1).

 \subsection{Radiative transfer}\label{radiative_transfer}
 
We use the 3-dimensional Monte Carlo radiative transfer
code {\tt RADMC} \citep{DD04a}, for which
a module to treat the emission from quantum-heated PAH molecules has been included. 
This module will be described in detail in Dullemond et al. (in prep.),
but a rough description has been given by Pontoppidan et al. (2006).
The code {\tt RADMC} \citep{DD04a, PON05} solves the temperature structure of the disk in a
Monte-Carlo way using a variant of the algorithm of Bjorkman \& Wood (1997).
This Monte-Carlo code also produces the source terms for scattering, in the
isotropic-scattering approximation.  

 


 \subsection{Dust model and PAH properties}

The dust is a mixture of grains in thermal equilibrium and transiently heated PAHs. 
\begin{itemize}
\item {\bf Thermal grains}: The grains are composed of graphite and silicate with optical constants from Draine (1985). They have a MRN \citep{MRN} size distribution ($n(a) \propto a^{-3.5}$) with a size between 0.01 $\mu$m and 0.3 $\mu$m. 
 Since no silicate emission features have been detected in HD97048 (see Sect. 2),
we have considered the hypothesis of thermally decoupling the
carbon and silicate grains (See Sect. 5.3). 
The optical to mid-IR opacity ratio of silicate is much smaller than that of carbon. When the grains are thermally decoupled,  the low temperature of the silicates produces very weak features and the emergent flux will be dominated by emission from the
carbon grains. But the absence of the silicate feature
could aslo be due to a geometrical effect, to a lower silicate
abundance, and/or larger silicate grains in the inner regions.
  \item {\bf Transiently heated grains} (PAH): 
  We can explain the observed PAH spectra, at least of the isolated HAe stars, with
PAH abundances and qualitative properties similar to those of PAHs in the ISM \citep{HAB04b}.
In the ISM, PAHs are made up of a few tens up to a few hundreds of carbon atoms; for reasons of simplicity, 
we take only one PAH size in our model, $N_C=100$. The hydrogen to carbon ratio is $H/C = f_H \times(6/N_C)^{0.5}$ (case of compact symmetric PAHs, see \citet{OMONT86}) with $f_H$ - the hydrogenation fraction of the molecule. Here, we consider $f_H$ equal to 1 (i.e., essentially fully hydrogenated PAHs) or 0.5 (partially hydrogenated PAHs). We take the absorption cross section from \citet{LID01} based on both laboratory data and astrophysical spectra. Those authors consider the bands at 3.3, 8.6, 11.3, 11.9 and 12.7 $\mu$m from vibrational modes of the aromatic C-H bond; the strong bands at 6.2 and 7.7 $\mu$m due to vibrations of the aromatic C-C bonds; and a few weak features probably caused by C-C bending modes at 16.4, 18.3, 21.2 and 23.1 $\mu$m. 
With respect to the charge, in order to keep the model simple, we
assume that all PAHs are neutral. Simple determination of the ratio between the photoionisation rate of the grain to the electron-grain recombination rate suggests that this is probably the case in the outer regions ($R\ge 150$ AU) of a disk heated by a typical HAe star (see \citet{HAB04a} for more details).
Finally, we do not take into account photo-destruction of PAHs in a strong FUV radiation field.
  
  \end{itemize}

In summary, we adopt a dust model with large thermal grains of graphite and silicate and small transiently heated aromatic particles. 
The silicate abundance in dust is $[Si/H]= 3.10^{-5}$, and the total carbon abundance in dust is $[C/H]= 2.2\textrm{\ }10^{-4}$. 
Of this 10\% are in PAH and 90\% in large grains.

\section{Comparison between model and observations}

In this section, we compare the model's predictions to the observations.
In the following, we first describe the adopted stellar and disk parameters for HD 97048
and then discuss the results of the calculation and the confrontation to the observations.

\subsection{Stellar and disk parameters}

A Kurucz model spectrum is taken for the
central star with T$_{eff}$ = 10000 K. 
The stellar parameters (L$_{\star}= 32 L_{\odot}$, M$_{\star}=2.5 M_{\odot}$) were chosen from stellar evolutionary tracks by Siess et al. (2000) for an age of 3 Myr \citep{LAGA06}.
Once the stellar flux has been redenned, the resulting SED is in agreement with the photometry extracted from \citet{HILL92} in the UV and near-IR. To correct for extinction, we used the method of \citet{CARDELLI} where we adopt A$_{V}$=1.24 \citep{VDA98} and E(B-V)=0.36 \citep{DAVIES,THE86}.


Concerning the disk's parameters, the surface density is taken equal to $\Sigma= \Sigma_{0}\textrm{\ }(r/R_{in})^{-q}$ with q the power law index equal to 3/2 inferred for the solar nebula \citep{WEID77} and $\Sigma_{0}= 444 \ g.cm^{-2}$ a minimum value deduced from VISIR observations \citep{LAGA06}. 
 The inner radius $R_{in}$ is at the dust (silicate) evaporation radius at 0.4 AU from the central star and the outer radius is at 370 AU as deduced from observations \citep{LAGA06}. 
Finally, as the disk is vertically optically thick at the wavelength of observation, a minimum mass of 0.01 M$_{\odot}$ (gaz + dust) can be derived \citep{LAGA06}. We take this minimum disk mass as a first guess.
 \subsection{Disk's structure}
\begin{figure}[!h]
 \centerline{\resizebox{8.55cm}{!}{\includegraphics[angle=90,scale=1.]{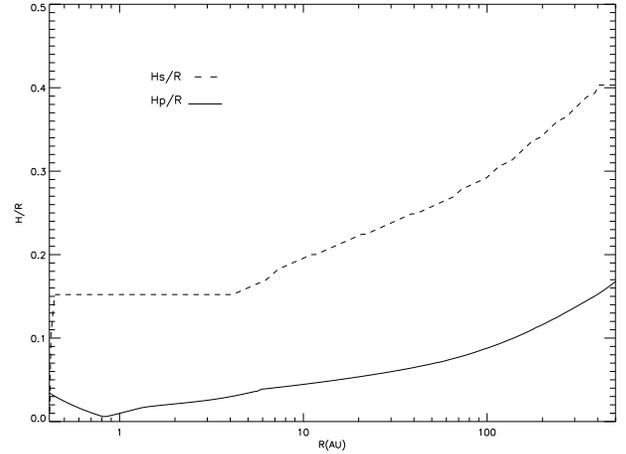}}}
 \caption{Structure of the disk for the template model. The surface (dashed line) and pressure (full line) scale height of the flared disk versus the distance from the central star.}\label{structure_template} 
 \end{figure}

 Figure~\ref{structure_template} shows the run with radius of the pressure ($H_{p}$) and photospheric ($H_{s}$) scale height. 
 The disk is flaring with H$_{s}$ increasing with radius as R$^{9/7}$ \citep{CG97}. At a radius of 135 AU, $H_{s}$= 51 AU.

In a previous work, we used the VISIR PAH band image to constrain the
parameters of the disk structure around HD 97048 \citep{LAGA06}.
Using a very simple model, \citet{LAGA06} measured a flaring index of 1.26$^{+0.05}_{-0.05}$ and $H_{s}$=51.3$^{+0.7}_{-3.3}$ AU at 135 AU.
Both values are very close to those expected from our hydrostatic, radiative equilibrium models of passive flared disks \citep{CG97,DDN01}.

Concerning the structure of the inner region, VISIR has clearly not enough spatial resolution to constrain it.
The puffed inner rim and the shadow region lie within the central pixel of VISIR (1 pixel is equivalent to 13.5 AU for a distance of 180 pc) and the two effects\footnote{Increasing the height of the inner rim augments the shadow region.} compensate each other in terms of resulting mid-IR emission.
Finally, we would like to underline that as soon as the disk is vertically optically thick in the mid-IR, the change of the slope of the surface density in the model does not have any influence on the structure of the disk or its mid-IR emission and cannot be constrained here.



\begin{figure}[!h]
 \centerline{\resizebox{8.55cm}{!}{\includegraphics[angle=90,scale=1.]{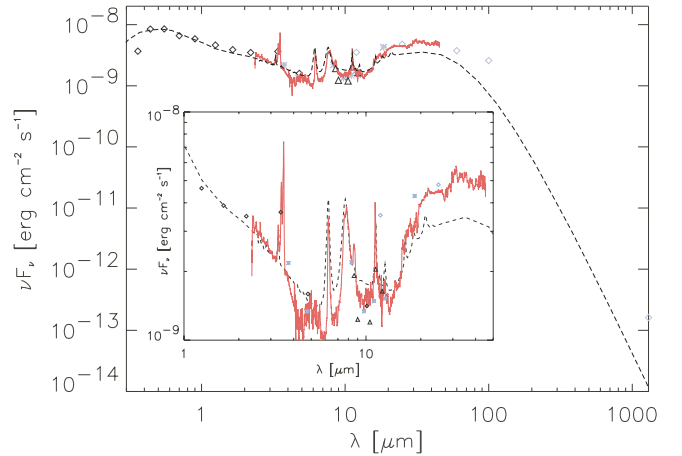}}}
 \caption{SED calculated with the template model (dashed line). The disk is inclined by 43 degrees and the system is situated at a distance of 180 pc. A Kurucz model spectrum is taken for the
central star with T$_{eff}$ = 10000 K. The luminosity is L$_{\star}= 32 L_{\odot}$ and the mass M$_{\star}=2.5 M_{\odot}$. The disk is flared with a total mass of 0.01 M$_{\odot}$, R$_{in}$=0.41 AU and R$_{out}$=370 AU. Full red line shows ISO SWS spectrum of HD~97048. Points of photometry are taken from \citet{HILL92} (open diamond black), IRAS (open diamond blue), \citet{PRUSTI94} (blue crosses) and VISIR measurement (black triangle)}\label{template_spectro}
 \end{figure}

  \begin{figure}[!h]
\centerline{\resizebox{8.55cm}{!}{\includegraphics[angle=0,scale=1.]{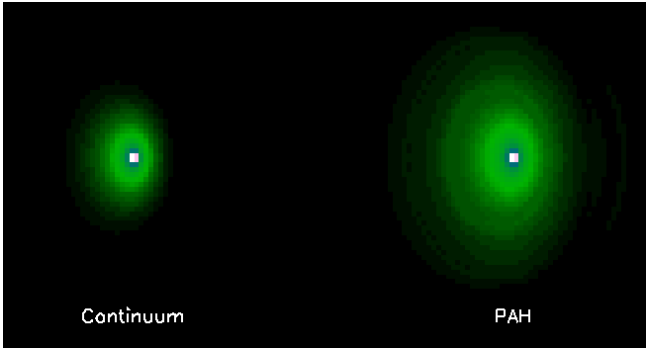}}}
 \caption{The non PSF-convolved predicted image for our template model in the adjacent continuum (at 10.5 $\mu$m on the left) and in the PAH band at 11.3 $\mu$m (on the right).}\label{template_ima}
 \end{figure}

\subsection{Spectral energy distribution}

Figure~\ref{template_spectro} shows the calculated spectrum in the 0.1-1300 $\mu$m range for a star/disk system
inclined by 43 degree and situated at 180 pc.
At short wavelength between 0.1 to $\sim$1$\mu$m, the emission comes from the stellar photosphere.
The near-IR emission (around 3 $\mu$m) is induced by the puffed inner rim. The region which emits between 5 and 8 $\mu$m corresponds to the shadow region. In this part, the disk is not in sight of the star itself, but it receives flux from the inner rim, which is sufficiently strong to keep the disk up. At larger radii, the flaring disk reappears from the shadow, and produces the rise observed in the spectrum around 20-30 $\mu$m.
Concerning the dust spectroscopic features, the PAH emission features are clearly visible, and some of them at 3.3, 6.2, 7.7 and 11.3 $\mu$m are very strong.
Strong ratio between 30 and 10 $\mu$m and PAH features show that the disk intercepts a large fraction of energy in the outer part and are indeed evidence of a flared disk geometry.
For geometrically flat disks, PAH features are predicted to be very weak, even when PAH with standard properties are present on the disk surface (Habart et al., 2004).
The infrared emission under the narrow PAH features is mostly due to the large thermal 
grains which are very hot in the inner regions. 
For the model in which carbon and silicate grains are thermally coupled, it is possible to see, for example, the strong 
broad feature due to silicate emission peaking at about 10 $\mu$m.
On the other hand, for the model in which the grains are decoupled, one can see (Fig.\ref{template_spectro}) that this feature almost disappears.

In Figure~\ref{template_spectro}, we also compare the predicted SED to
one of HD~97048 constructed with different sets of data.
Photometric points from the visible to the mid-IR were taken from \citet{HILL92} and \citet{PRUSTI94}. To that we have added IRAS and VISIR photometric measurements. Spectroscopic observations were obtained with ISO-SWS (Short Wavelength Spectrometer) \citep{ACKE04b}. 
 The model reproduces correctly the global shape of the
observed SED of HD97048. 
The agreement between predicted slopes and absolute fluxes from the near- to mid-IR waves with observed ones is rather acceptable (differences $\le$20-40\%).
 Moreover, our model reproduces also well the observed intensity of the most commonly observed PAH features, i.e., 3.3, 6.2, 7.7 and 11.3 $\mu$m, especially if we correct for contamination by the associated reflection nebula (about 40\%, see Section 3).  
This is in agreement with previous comparisons made by Habart et al. (2004)
between flared disk model results and observed PAH emission features with ISO and ground-based telescopes of some thirty HAeBe stars, including HD~97048.

Nevertheless, one can note that there is some mismatch between the predicted and observed PAH spectra;
 the strength of the 8.6 $\mu$m feature is, for example, significantly underestimated by the model. 
However, this is not surprising considering the uncertainties on the PAH absorption cross section (Li \& Draine 2001) 
and our simple hypothesis that PAH are characterized by a single charge state, size or hydrogenation parameter along the disk.
This is unlikely to be the case and is briefly discussed in this section.
Also, one can note that HD 97048 presents peculiar strong features that peaks at 3.43 and 3.53 $\mu$m
 not predicted by our model. 
Several studies have proposed attributing these features to surface C-H stretching modes on  nanodiamond particles
\cite[e.g.,][]{GUILL99,VKK02,SHEU02,JONES04}.
Because of the good match between laboratory and observed spectra, this identification appears convincing.

In addition, the absence of the small grains of silicate features in the HD 97048 spectra could appear intriguing since they are the most abundant dust species in interstellar space.
However, in ISO spectra, Acke \& van den Ancker (2004) reported non detection of silicate feature for 16 objects out of 46 
showing that the absence of the silicate feature is a common phenomenon among HAeBe stars. 
The silicate emission at 10 $\mu$m is arising from grains with a size of 0.1 $\mu$m thermally heated in the inner region (1 $<$ r $<$ 10 AU) from direct and/or indirect irradiation by the central star. The absence of this emission could probably result from various effects concerning either the dust properties either the disk geometry in the inner region. Here, we have considered the thermal decoupling hypothesis but it
could also be to less silicate abundance in the inner disk part.
Decreasing globally the silicate abundance could not be a solution
since the 20 microns feature shows the presence of silicates
in the 50 AU $<$ R $<$ 100 AU parts of the disk. We have also tested different structures of disk "made by hand" and we have shown 
that it is possible to explain the abscence of the silcate emission at 10 µm 
with a geometrical effect. But at the moment, the structure is calculated 
SELF-CONSISTENTLY and the aim of the paper is not 
to construct a structure by hand that could fit all the data of HD97048 but 
won't be physical. Furthermore, this issue could not be invistigated with the resolution of VISIR observations. 

Finally, concerning the longer wavelength, we find that the submm flux is too low by a factor 70 compared to observations \citep{HENN98}. This shows that a large reservoir of large grains (around a millimeter size) must exist in the outer regions of the disk. These grains would naturally reside close to the midplane and therefore do not affect the shape of the disk. They are therefore not within the focus of this paper, and for this reason we do not include them in our model. If we would have included them, they would only affect these long-wavelength fluxes, because the disk has a flaring geometry (see e.g.~Dullemond \& Dominik 2004a, Figs.\ 6 and 7). 

 \begin{figure*}[t!]
\begin{center}
\begin{minipage}[t]{9cm}
 \centerline{\resizebox{8.8cm}{!}{\includegraphics[angle=90,scale=1.4]{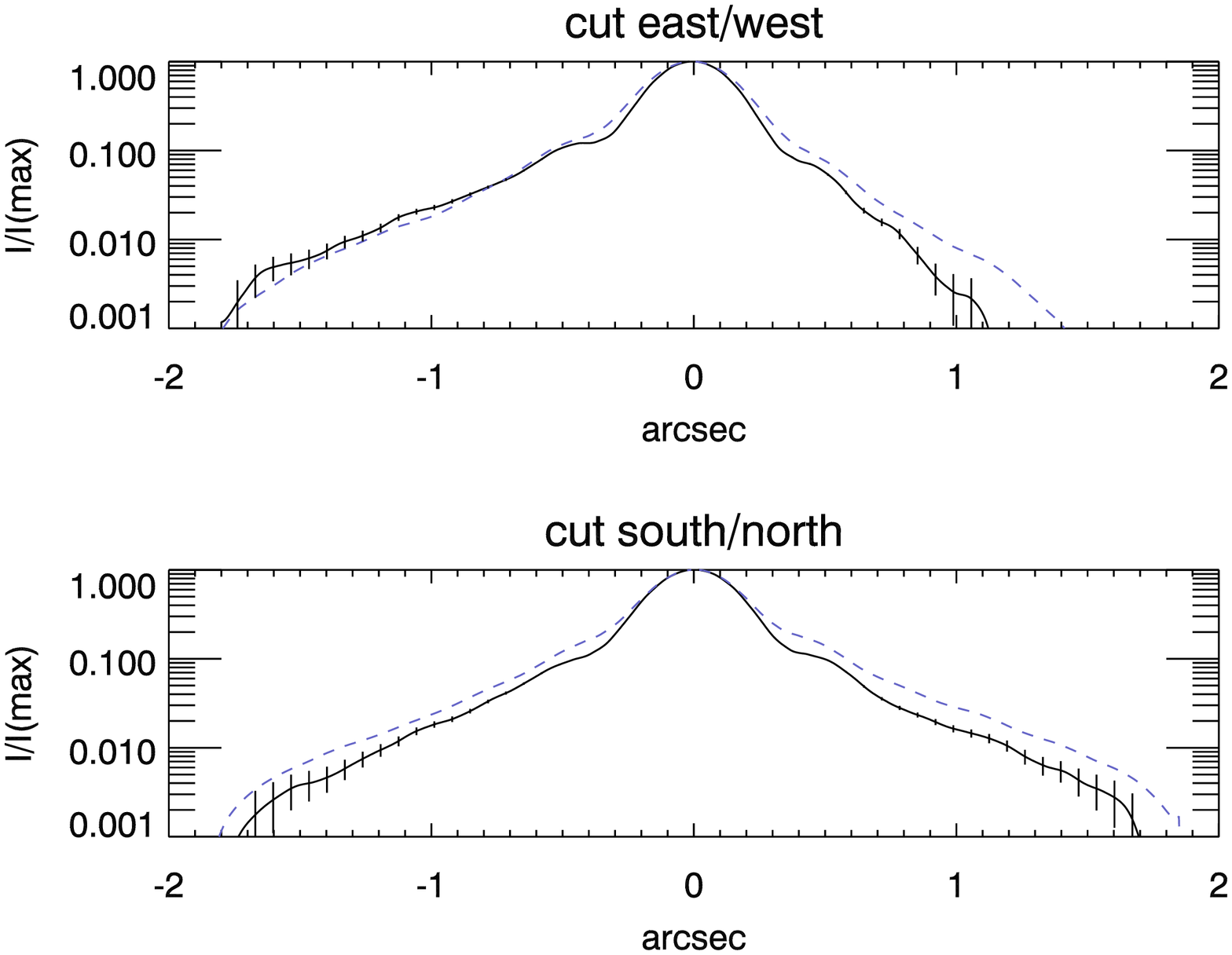}}}
\end{minipage}%
\begin{minipage}[t]{9cm}
\centerline{\resizebox{8.8cm}{!}{\includegraphics[angle=90,scale=1.4]{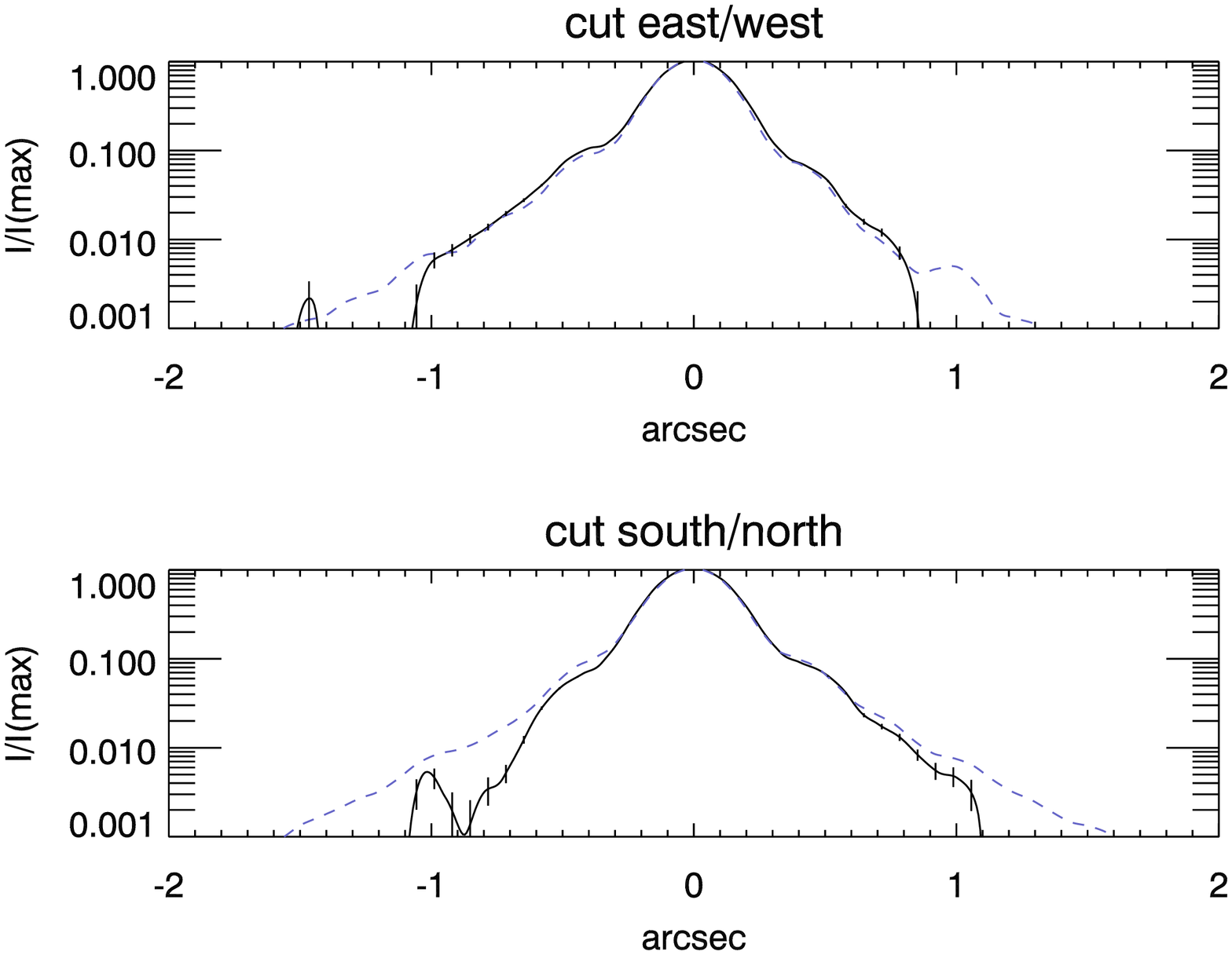}}}
\end{minipage}
 \caption{\emph{On the left}, cut in north/south and east/west (up panel) for the template model of the PAH band at 11.3 $\mu$m convolved with VISIR PSF (dashed line) compared to the observation (full line). \emph{On the right}, the same as on the left with the adjacent continuum at 10.5 $\mu$m. The errors drawn are 1-$\sigma$ errors RMS.}\label{comparaison_model_obs}
 \end{center}
 \end{figure*}

\subsection{Imaging}

Figure~\ref{template_ima} shows the modelled disk image in the PAH band at 11.3 $\mu$m and the adjacent continuum. These emissions both originate from the optically thin surface of the disk but the PAH emission is much more extended than the adjacent continuum. 
The continuum reaches 50 \% of its integrated intensity at a very small radius (about 2 AU), while the PAH feature does so at large radii (about 100 AU). This behaviour basically reflects the different excitation mechanisms of grains in thermal equilibrium and PAHs transiently heated. Only grains very near the star are warm enough to emit at 11 $\mu$m whereas PAHs farther away can be excited and emit in the 11.3 $\mu$m feature. The emission at 11.3 $\mu$m (band + continuum), in the central part of the disk ($<$ 2 AU) is dominated by the thermal emission from silicate and carbon grains, whereas in the outer regions of the disk, the emission is dominated by the PAH feature. 
Morever, among the PAH features, the 11.3~$\mu$m is one of the most extended spatially. Indeed, as discussed in \citet{HAB04a},
 the features at shorter wavelengths are in fact stronger in
the inner part of the disk (where PAHs are hotter because multiphoton events occur where the radiation field is most intense), decreasing rapidly in the outer cold part.
On the other hand, the features at longer wavelengths are more extended. 

 Fig~\ref{comparaison_model_obs} shows the brightness emission profiles of the PAH feature and the adjacent continuum 
obtained by convolving the model with the corresponding PSF observed by VISIR.
The model predicts a spatial distribution of the PAH emission very similar to that observed. 
The predicted FWHM and wings extension are in fact very closed to the observed ones (differences $\le$20\%).
Moreover, our disk model predicts, as in the observations, an asymetry east/west, which only results from the inclination of a flared disk optically thick at the observed wavelength. 
This acceptable agreement gives a strong support for the physics underlying in our flared disk model.
The disk parameter that most affects the PAH emission is in fact the disk flaring angle, 
which determines at each radius the fraction of FUV intercepted by the disk surface.
Lower values of the flaring could be caused by a variety of reasons, for
example if the dust settles toward the disk midplane \citep{DD04b}.
If the disk is less flared, the PAH emission which directly tracks the illumination of the disk surface
 will strongly reduce in the outer disk region. 
Less flared disks will have less extended PAH emission features and weaker.
This will be particularly true for the features at long wavelength,
such as the 11.3 $\mu$m one, which have a large contribution from the outer disk. 
In the extreme case of a fully self-shadowed disks, 
the PAH feature strenghts should decrease by orders of magnitude and the spatial distribution should be similar
to one of the adjacent continuum.

In addition, it is remarkable to note that, as predicted by the model,
 the observed spatial extension of the 11.3~$\mu$m feature
is much larger than that observed for PAH feature at the 3.3 $\mu$m, which are   
extended on a scale of (several) 10 AU \citep{HAB04b}.
This has the interesting implication that PAHs appears to be present 
 over a large range of radius; in other words, PAHs can survive over a wide range of physical conditions.
Finally, concerning the spatial extent of the 10 $\mu$m adjacent continuum emission, the model
 predicts that it is slighlty broader than the PSF but still agrees with the observations.

It must be emphasized here that there are several complications which we have neglected. 
The most obvious is that we have assumed that PAHs can be characterized by a single size, hydrogenation and charge state. 
This is unlikely to be the case, and one can expect variations as a function of radius and depth in the disk.
For example, PAHs are likely to be more positively ionized in the inner disk regions. 
Moreover, processes such as photo-evaporation or coagulation could affect the abundance and size of PAHs.
In order to get some insight into the specific
PAH properties, one needs spatial information of several band strength ratios. 
We are developing this study in a forthcoming paper.

.
 \section{Summary and conclusions}\label{summary}
 In a former paper \citep{LAGA06}, we were able to constrain for the first time the flaring geometry of a disk around an intermediate-mass young star HD97048. These results were based on a very simple model making several assumptions such as a surfacic PAH emission, an optically thick disk, and a power-law function for the surface height and the intensity. In the present paper, using a full radiative transfer model based on predicted disk geometries assuming hydrostatic equilibrium \citep{DDN01, DD03, DD04a}, we could:
 \begin{itemize}
 \item justify the hypothesis made in the previous paper \citep{LAGA06}, and therefore confirm the results, e.g., the images calculated with a different code \citep{PINTE06}.
 \item show that both SED and spatial distribution of the PAH emission and the adjacent continuum are in very good agreement with the model predictions for common disk parameters.
 \item take the general agreement between observations and predictions as a strong support for the physical pictures underlying our flared disk model.
 \end{itemize}

 \bibliographystyle{aa}
\bibliography{biblio_hd97048}

 \end{document}